\documentclass{mn2e}
\usepackage{amssymb}
\usepackage{ulem}
\usepackage{color}
\input epsf.sty
\newif\ifAMStwofonts
\AMStwofontstrue

\begin{document}

\title[Black hole spin in SDSS J094533.99+100950.1]{Constraints on the black hole spin in the quasar SDSS J094533.99+100950.1}

\author[Czerny et al.]{B. Czerny$^1$\thanks{bcz@camk.edu.pl}, K. Hryniewicz$^1$\thanks{krhr@camk.edu.pl}, M. Niko\l ajuk$^{2,3}$\thanks{mrk@alpha.uwb.edu.pl}, and A. S\c adowski$^1$\thanks{as@camk.edu.pl}\\
$^1$ Copernicus Astronomical Center,
ul.~Bartycka~18, P-00\,716~Warszawa, Poland\\
$^2$ INTEGRAL Science Data Centre, Chemin d'Ecogia 16, 1290 Versoix,
Switzerland\\
$^3$ Faculty of Physics, University of Bia\l ystok, Lipowa 41, 15-424 Bia\l ystok, Poland}

\maketitle

\begin{abstract}
The spin of the black hole is an important parameter which may be responsible for the properties 
of the inflow and outflow of the material surrounding a black hole. 
Broad band IR/optical/UV spectrum of the quasar SDSS J094533.99+100950.1 is 
clearly disk-dominated, with the spectrum peaking up in the observed frequency range.
Therefore, disk fitting method usually used for Galactic black holes can be
used in this object to determine the black hole spin. We develop the numerical code for
computing disk properties, including radius-dependent hardening factor, and we apply the ray-tracing method
to incorporate all general relativity effects in light propagation. 
We show that the simple multicolor disk 
model gives a good fit, without any other component required, and the disk extends down to the marginally 
stable orbit.
The best fit accretion rate is 0.13, well below the Eddington limit, and the black hole spin is moderate, 0.3.
The contour error for the fit combined with the constraints for the black hole mass and the disk inclination
gives a constraint that the spin is lower than 0.8. We discuss the sources of possible systematic errors
in the parameter determinations.

\end{abstract}



\section{Introduction}

The role of the black hole spin in the accretion process and in the jet formation is not yet well understood.
Spin paradigm connect the high values of the black hole spin with the presence of fast, well collimated jets
(e.g. Sikora, Stawarz \& Lasota 2007). Spin also affects the overall accretion efficiency which may be important from the point of view of
the global evolution of AGN and their feedback to host galaxies.

There are five known methods of black hole spin determination: (i) fitting the shape of the broad iron 
K$\alpha$ line to X-ray data  {\bf (ii) connecting the observed variability timescale to the innermost stable orbit (iii) estimating
the accretion efficiency (iv) analysis of the orbital motion in the case of binary black hole (v) fitting the accretion disk continuum to the broad band data.}

The first method has been used to a number of AGN, mostly nearby Seyfert galaxies with high quality 
X-ray spectra (MCG-6-30-15: $a=0.989^{+0.009}_{-0.002}$, Brenneman \& Reynolds 2006; $ a > 0.8$, Goosmann et al. 2006; $0.86^{+0.01}_{−0.02}$, Perez et al. 2010; MRK 509: $0.78^{+0.03}_{−0.04}$, Perez et al. 2010; 
Fairall 9: $a = 0.60 \pm 0.07$, Perez et al. 2010; modest preference for a rotating black hole seen in MCG –5-23-16, Ark 120, NGC 2992, and 3C 273 by Brenneman \& Reynolds 2009; also combined spectra for a quasar sample
in the  Lockman Hole indicated preference for a spinning black hole, as argued by Streblyanska et al. 2005).

The second method is based in the independent estimates of the mass and the variability timescale through the assumption that this timescale is set by the 
dynamical timescale at ISCO orbit (Abramowicz \& Nobili 1982). It was applied to several gamma-ray loud blazars by Xie et al. (2003), and the sources with black hole masses larger than $2 \times 10^8 M_{\odot}$ in their sample of 17 objects required a spinning black hole with $a > 0.6$. 

The third method is applied in the statistical approach to whole quasar populations. The estimates done by Soltan (1982) indicated the overall accretion efficiency of order of 0.1, corresponding to a moderate black hole spin, and the subsequent papers based on the same approach came with similar values (e.g. 0.07 - 0.15, Elvis et al. 2002; $0.06^{+0.026}_{-0.013}$, Martinez-Sansigre \& Taylor 2009; 0.065-0.075, Shankar et al.2009). Shankar et al. (2010)
obtained the efficiency higher than 0.17 (corresponding to the Kerr parameter greater than 0.92) at the basis of the modelling the quasar luminosity function in the redshift span 3 to 6.

The fourth method was applied to the source OJ 287 which is the strongest and the best studied candidate for 
a binary black hole system. The analysis of the binary motion indicates the spin of the more massive component  was estimated to be $0.28 \pm 0.01$.   

The fifth method was applied to the analysis of the Galactic sources (e.g. Shafee et al. 2006, McClintock et al. 2006, 
Middleton et al. 2006, Gou et al. 2010) but not to single AGN objects, since it requires the broad band spectrum to be the disk-dominated, and the spectral region corresponding to
the maximum of the disk emission has to be clearly visible. This second condition is particularly difficult to  meet in AGN since the spectrum usually peaks in the unobserved far-UV/soft X-ray band. We apply this method to an active nucleus SDSS J094533.99+100950.1 which satisfies such requirements.

\section{Data and numerical models}

\subsection{data}
\label{sect:data}

SDSS J094533.99+100950.1 {\bf ($z= 1.66$)}  belongs to the class of Weak Line Quasars (McDowell et al. 1995, Fan et al. 1999, Reimers et al. 2005, 
Shemmer et al. 2006, Leigly et al. 2007, Diamond-Stanic et al. 2009, Plotkin et al. 2010) but the broad Mg II line
is well visible in this source which allows to measure the black hole mass (Hryniewicz et al. 2010).

\begin{figure}
\epsfxsize=8cm
\epsfbox{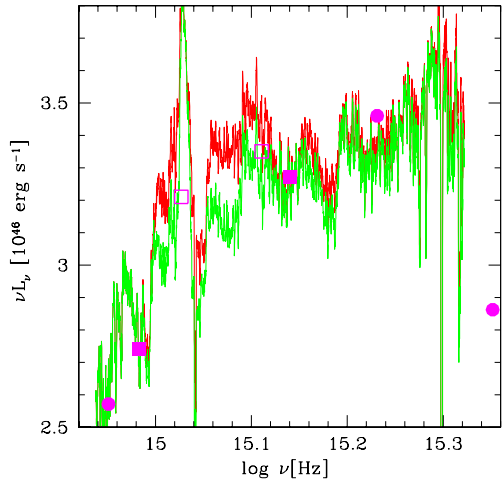}
\caption{The UV continuum of the SDSS J094533.99+100950.1 spectrum (blue continuous line) 
dereddened for the Galactic reddening, the same continuum with an exemplary subtraction of the iron
contamination using the templates of  Vestergaard \& Wilkes (2001). Large filled dots mark the original SDSS photometric 
points used in our analysis, large open squares mark the removed SDSS photometric points $r$ and $i$, 
highly contaminated by Fe II and Mg II line and filled squares mark the additional two points 
used to replace $r$ and $i$.
}
\label{fig:photometry_spect}
\end{figure}

We combine the broad band photometry given by Hryniewicz et al. (2010) with
the spectrophotometry from SDSS data base (Adelman-McCarthy et al. 2007). We take the photometric points from GALEX (Martin et al. 2005) and
2 MASS survey (Skrutskie et al. 2006), and three photometric points out of five available from SDSS. 
The other two SDSS photometric points ($r$ and $i$) are strongly contaminated
by  Mg II emission line, and FeII emission. The problem is illustrated in 
Fig.~\ref{fig:photometry_spect}. 
Therefore we choose two other
spectral windows, the best from the point of view of FeII emission, consistent with
the windows suggested by Vestergaard \& Wilkes (2001). We calculated the flux there from the 
spectroscopic data, taking the window width $\Delta \log \nu$ equal 0.002. 
The final set used throughout the paper
is given in Table~\ref{tab:tab1}, and plotted in Fig.~\ref{fig:best_fit}. We see that
broad band spectrum is well covered, in particular due to near-UV and far-UV points from GALEX.

These data points are based on assumption that there is no intrinsic reddening of the source.
However, circumnuclear dust in the form of dusty/molecular torus is present in AGN, even in bright 
quasars, and in some cases this dust may partially lie along the line of sight to the nucleus.
We address this issue in the Discussion. 

{\bf There is no evidence for a strong emission in the X-ray band. No X-ray data (XMM–Newton, RXTE, INTEGRAL) were found by Hryniewicz et al. (2010), and no source with a flux significantly higher than the X-ray background is found at its position in the ROSAT survey (Soltan, private communication).}

\begin{table}
\caption{The photometric points used for fit the data.}
\label{tab:tab1}
\begin{center}
\begin{tabular}[t]{ccccccccc}
\hline

Band & $log\nu$ & $log\nu L_{\nu}$ \\
     &   [Hz]   &  erg s$^{-1}$    \\
\hline
\\
GALEX FUV   & 15.721 & 45.563 $\pm 0.057$ \\
GALEX NUV   & 15.547 & 46.214 $\pm 0.040$ \\
 u & 15.350 & 46.450 $\pm 0.014$ \\  
 g & 15.228 & 46.523 $\pm 0.011$\\  
   & 15.140 & 46.515 $\pm 0.018$\\  
   & 14.983 & 46.438 $\pm 0.031$\\                       
 z & 14.950 & 46.393 $\pm 0.014$\\  
2MASS J   & 14.804 & 46.257 $\pm 0.064$\\
2MASS H    & 14.684 & 46.113 $\pm 0.085$\\
2MASS K$_s$   & 14.569 & 46.032 $\pm 0.099$\\
\hline
\end{tabular}
\end{center}
\end{table}

\subsection{disk structure and spectra models}

\subsubsection{disk structure}

As our basic model of the disk structure, we adopt the classical solution of Novikov \& Thorne (1973)
which generalized the Shakura \& Sunyaev (1973) model to include general relativity effects. If the 
data fits require accretion rates close to the Eddington value, we use more advanced disk structure
table model.  

The Novikov-Thorne model
is based on assumption of  
zero torque inner boundary condition at the ISCO (Innermost Stable Circular Orbit),
and using it we also assume that the disk thickness is small so the photons are emitted from the 
equatorial plane (see the discussion of this assumption by S\c adowski et al. 2009). 
In this case the local disk emissivity is not sensitive to any assumptions about the disk vertical
structure, including viscosity, and the local disk emissivity is given by the dissipation in the Keplerian disk in the
Kerr metric (Novikov \& Thorne 1973; Page \& Thorne 1974, Zhang et al. 1997). 
The disk model is fully described by the black hole mass, $M$,
and the accretion rate. We use  the dimensionless 
accretion rate, $\dot m$, expressed in units of the Eddington
accretion rate with efficiency appropriate for a Schwarzschild black hole:
\begin{equation}
\dot M_{Edd} = {64 \pi G M \over c \, \kappa_{es},} 
\end{equation}
i.e. the actual flow efficiency is not included in this scaling. The advantage of the model is that
the computations are fast and can be done with arbitrary accuracy in accretion rate and 
Kerr parameter, $a$. We allow both for a positive and negative black hole spin.

The propagation of radiation is done through ray tracing in the Kerr
metric (Cunningham 1975), and in the description we follow the recent papers of Niedzwiecki (2005), 
\. Zycki \& Niedzwiecki (2005), Svoboda et al. (2009) and Yuan et al.
(2009) (there are some misprints in the 
transformations given in the Appendix of the last work). We do not calculate the maps seen by the distant observer,
so we do the ray tracing from the disk to infinity, storing the results in various inclination bins.
The best fit is thus given by $M$, $a$, $\dot m$ and $\cos i$, where $i$ is the favored 
inclination angle of an observer.

\subsubsection{hardening factor}
\label{sect:hardening}

\begin{figure}
\epsfxsize=8.5cm
\epsfbox{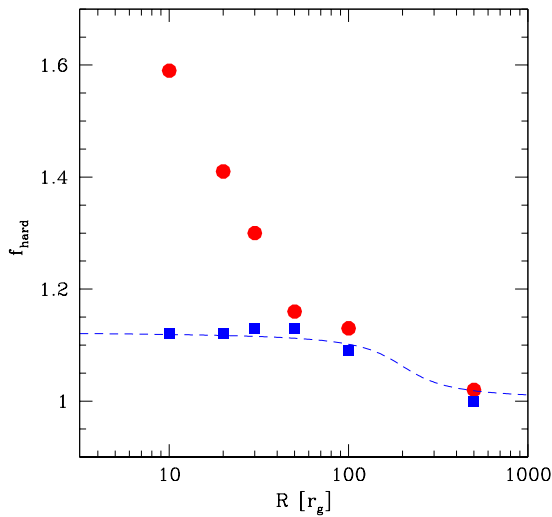}
\caption{The hardening factor as a function of radius for the accretion rate $\dot m = 0.3$, viscosity parameter $\alpha = 0.1$,   and
two values of the black hole mass: $10 M_{\odot}$ (dots) and $2.7 \times 10^9 M_{\odot}$ (squares). 
}
\label{fig:hardening_factor}
\end{figure}

Models of the radiative transfer in the disk atmosphere indicated the departure from the black body
(e.g. Shimura \& Takahara 1995). In the simplest approach this effect can be modeled assuming a constant hardening factor,
i.e. replacing the local disk effective temperature with the color temperature higher by a factor 
$f_{hard}$, with the total radiation flux unchanged. However, the IR part of the spectrum comes from larger radii where the disk atmosphere is less dominated by the electron scattering so the radius-independent hardening factor is unlikely to be a good approximation. 
There were some attempts of developing proper description of the spectra. Davis et al. (2007) applied the  hydrogen-helium
models of Hubeny et al. (2000) in a statistical study of quasar slopes in SDSS sample. However, the negligence of the metals is not acceptable
in case of relatively cool AGN disks.  Complete  models with numerous bound-free transitions were developed by
Davis et al. (2005) and the models based on this approach were even implemented in {\sc xpec} and used
to fit the spin in black hole binaries (see Done \& Davis 2008,  Kubota et al. 2010, Gou et al. 2010 and the references therein) but the tables do not cover 
supermassive black holes.

Therefore we estimated the hardening factor as a function of the disk radius
 by performing computations of the disk vertical structure using the code described in detail by Rozanska et al. (1999).
We calculated the vertical profile of the disk temperature, $T(z)$, the total (i.e. absorption plus scattering) Kramer's opacity, $\tau (z)$ and the effective optical depth, 
$\tau^*(z)$,
and we estimated the hardening factor in the following way:
\begin{equation}
f_{hard} = {T(\tau^* = 2/3) \over T(\tau = 2/3) }
\end{equation}
for a number of disk radii. We see that the hardening factor is about 1.125 in most part of the disk (see Fig.~\ref{fig:hardening_factor}), decreasing to 1 above $\sim 200 R_g$. Its value is much lower than expected in Galactic sources since the disk temperature is lower and the level of ionization is low enough to
allow efficient thermalization.  This radial dependence is fitted numerically and applied to calculated spectra models. For the black hole mass $2.7 \times 10^9 M_{\odot}$ a good simple fit can be used for all models:
\begin{eqnarray}
f_{hard} = 1 + A - {2A \over \pi} atan{(\log R - \log B)*20}\nonumber \\
A = 0.5*(1.125({\dot m \over 0.1})^{0.028} - 1) \nonumber  \\
B = 55 ({\dot m \over 0.1})^{1/3}.
\end{eqnarray}

\section{Results}

\label{sect:results}

The broad band photometry of SDSS J094533.99+100950.1 has been fitted with the Novikov-Thorne disk model, taking
into account the correction to the local black body emission in the form of the radius-dependent hardening 
factor described in Sect. ~\ref{sect:hardening}. The  10 photometric data points were fitted with the
models with four free parameters: the black hole mass, $M$, dimensionless accretion rate, $\dot m$, Kerr parameter, 
$a$ and the cosine of the inclination angle, $\cos i$. The Kerr parameter was allowed to take both positive and 
negative values. The inner disk radius was fixed at the ISCO appropriate for a given Kerr parameter. 
Quoted $\chi^2$ is thus given for six degrees of freedom, and the errors give 90 \% confidence levels.

\begin{figure*}
\epsfxsize=8.5cm\epsfbox{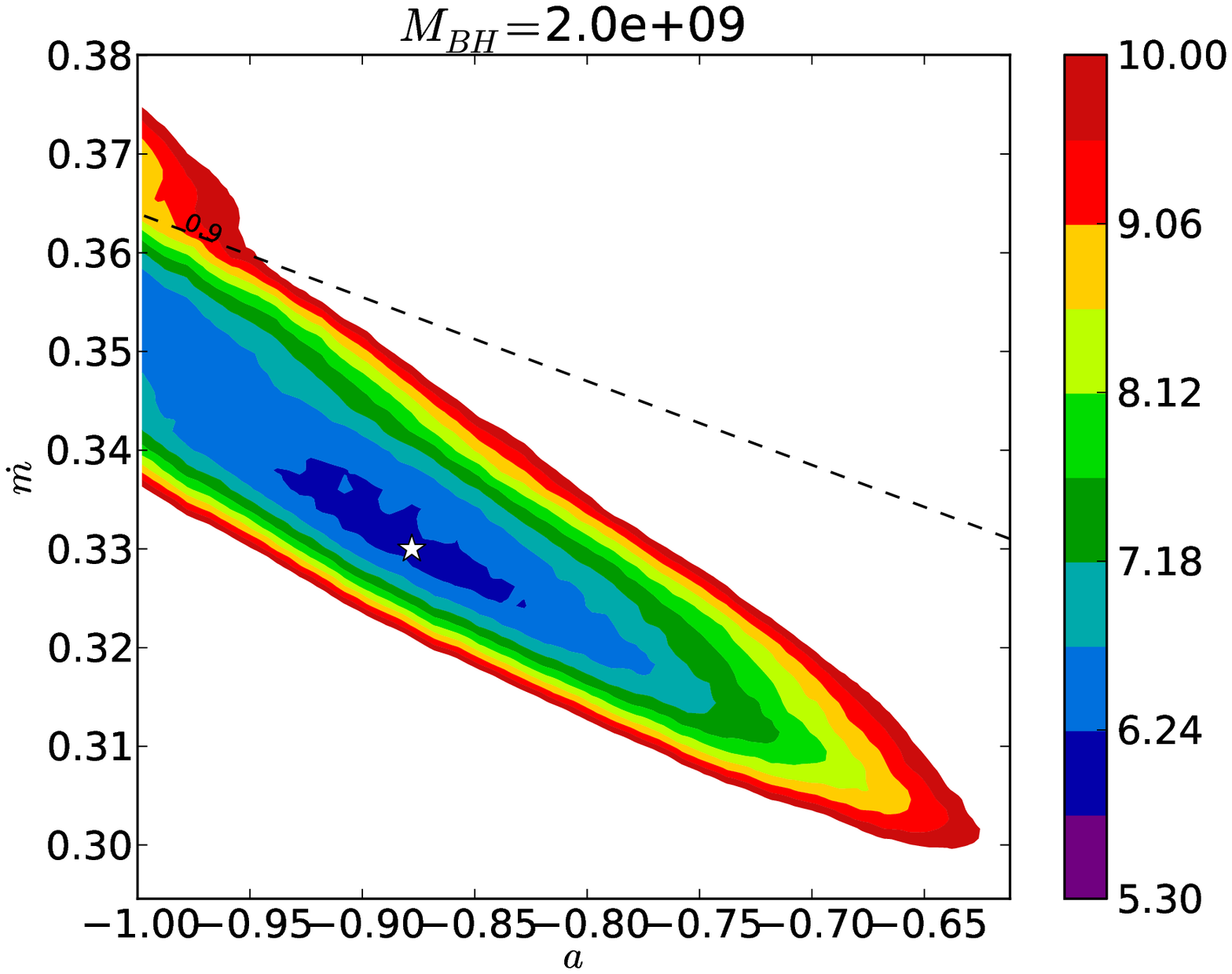}
\epsfxsize=8.5cm\epsfbox{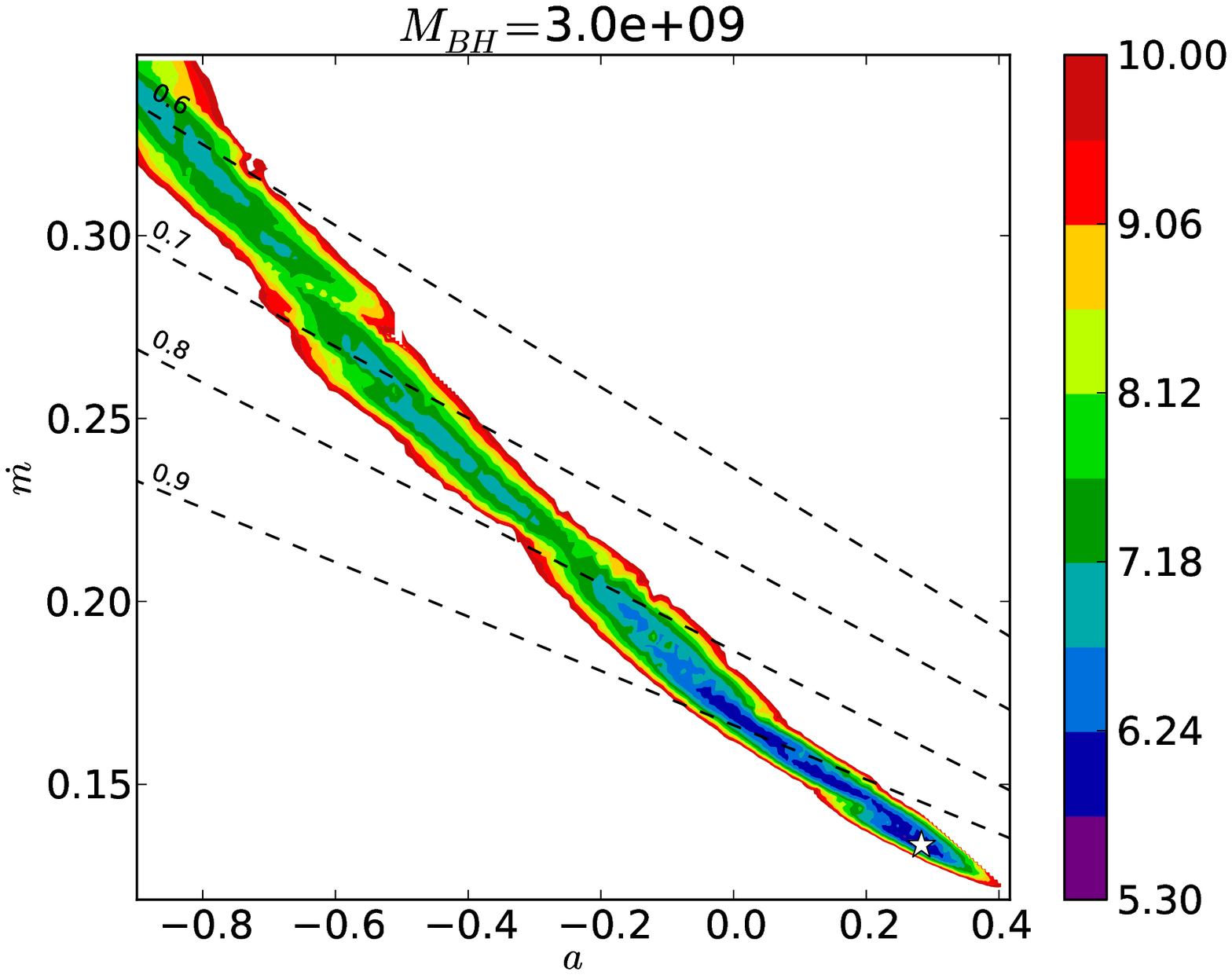}
\epsfxsize=8.5cm\epsfbox{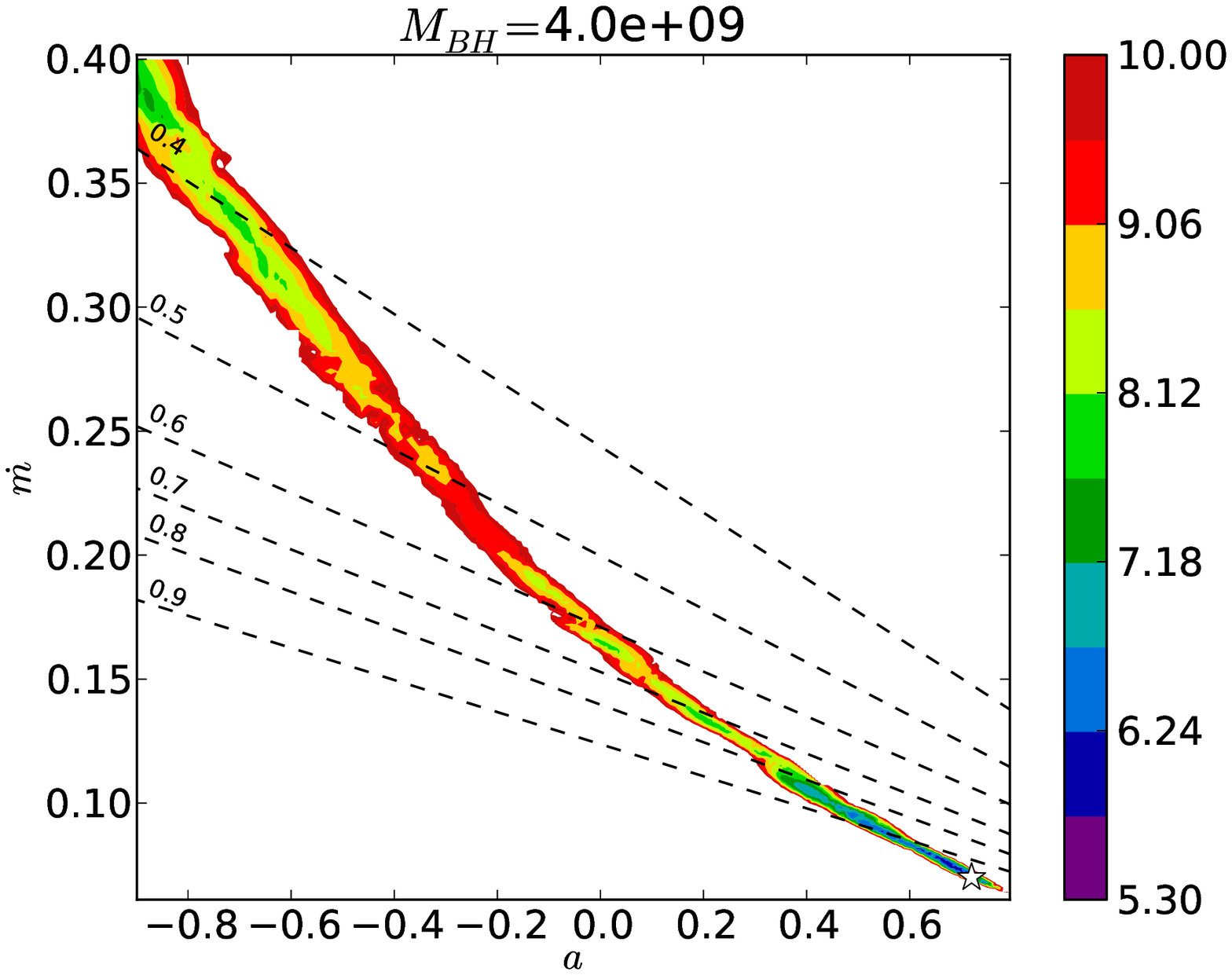}
\epsfxsize=8.5cm\epsfbox{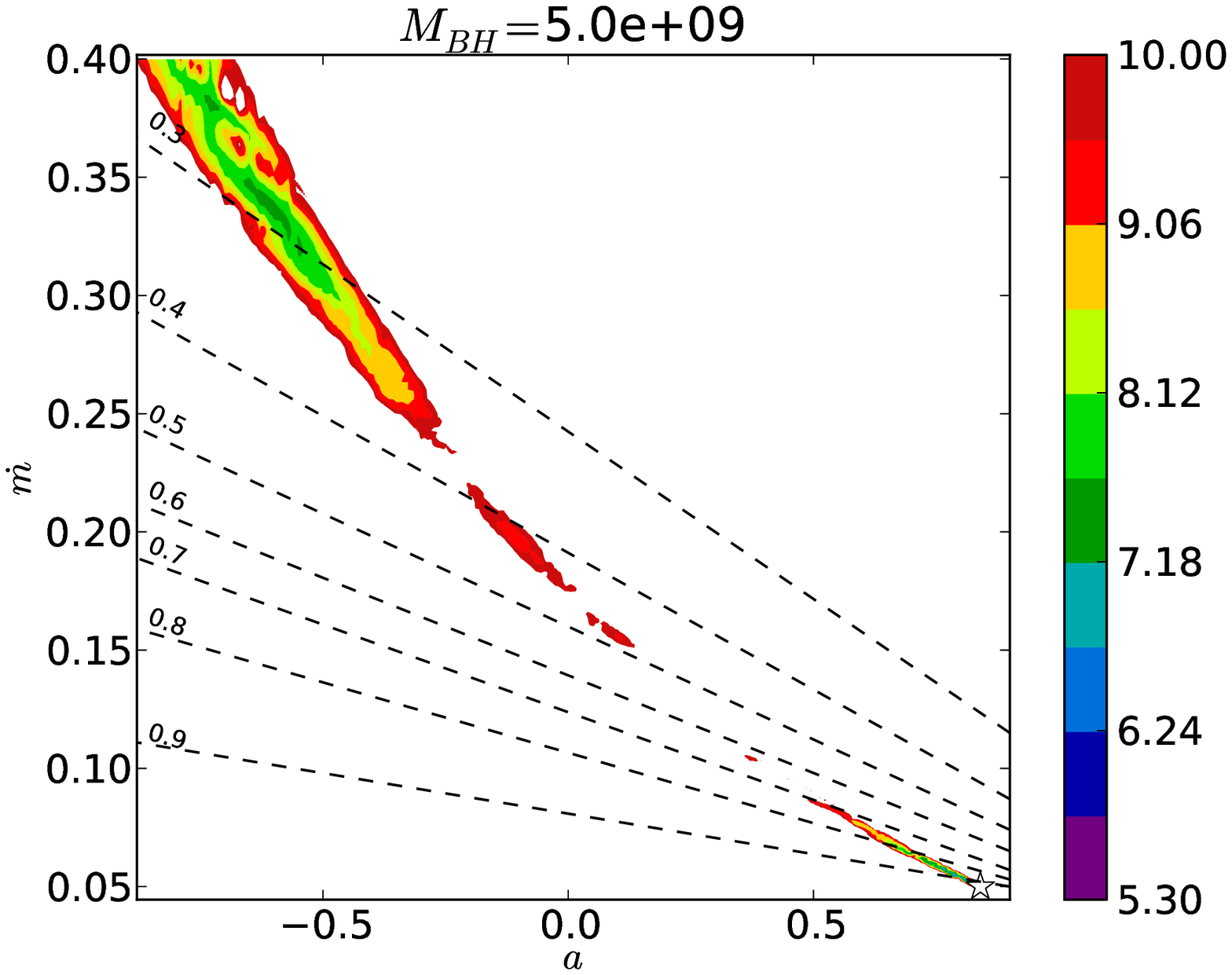}
\epsfxsize=8.5cm\epsfbox{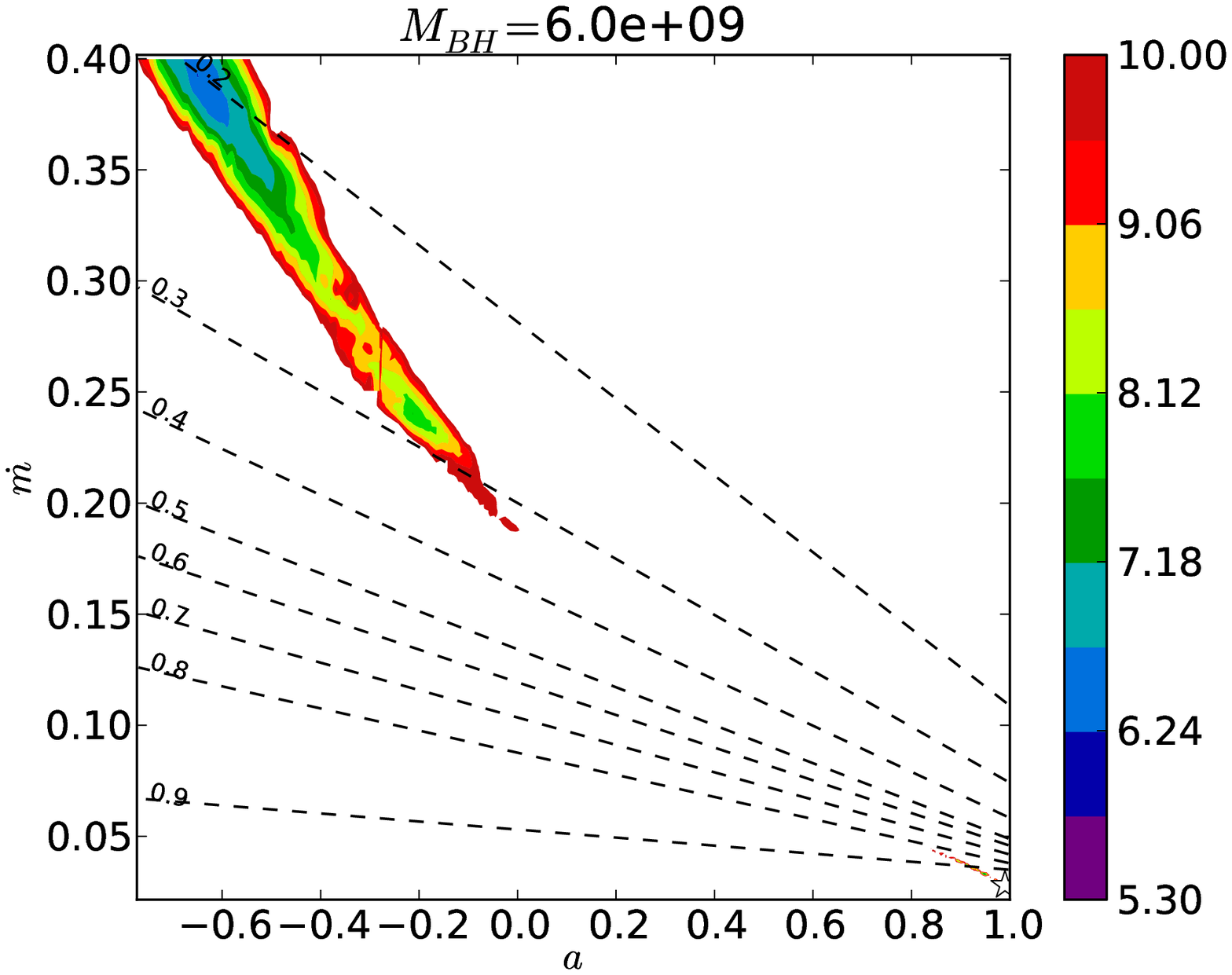}
\epsfxsize=8.5cm\epsfbox{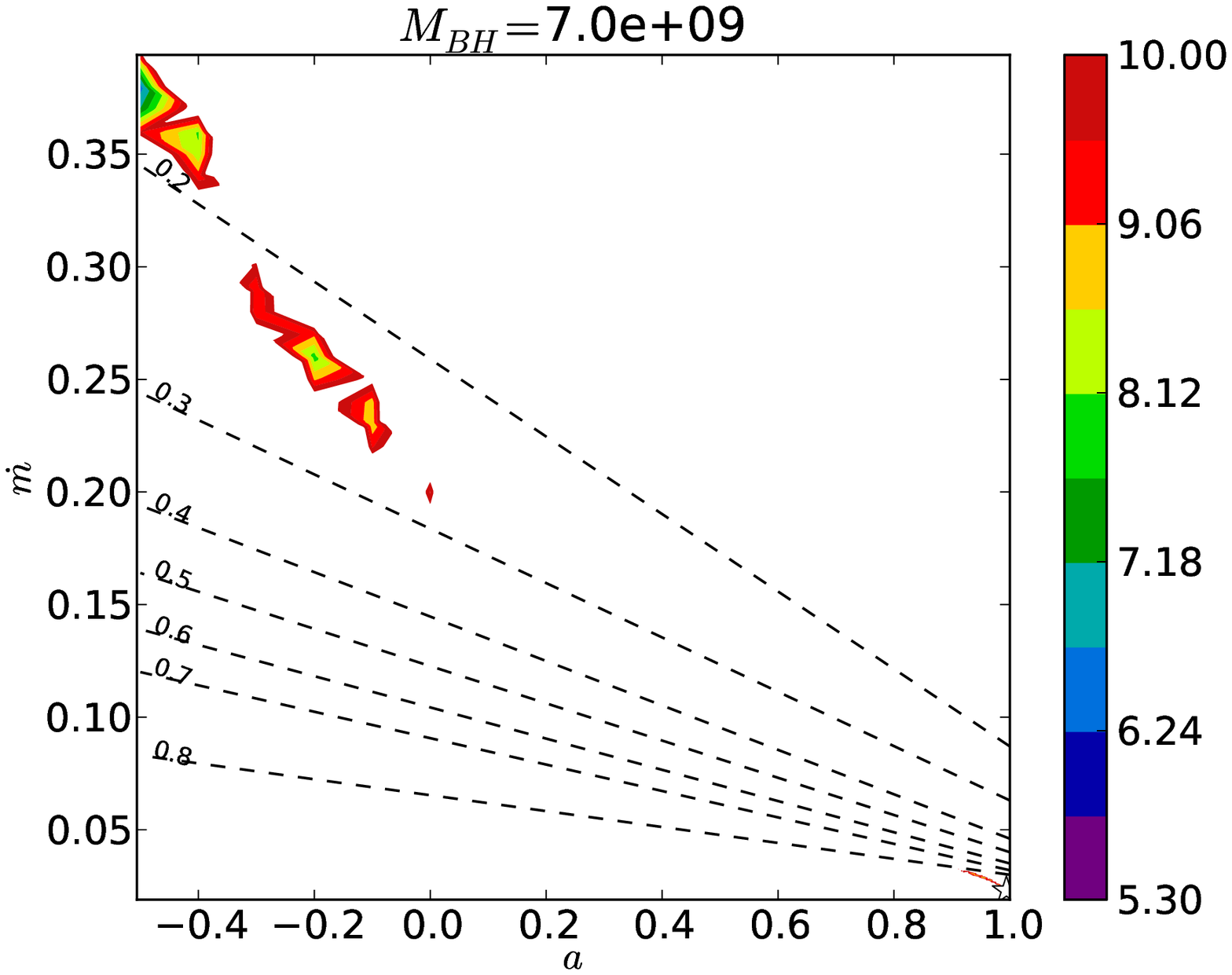}
\caption{The contour errors for the Kerr parameter, $a$, and the accretion rate, $\dot m$, for the black hole 
for the range of the black
hole masses from $2 \times 10^{9} M_{\odot}$ (top) to $7 \times 10^{9} M_{\odot}$ (bottom). The star indicates the position of the best fit solution, and dashed lines show the best fit values of the cosine of the disk inclination. 
For larger black hole masses two separate
best fit regions exist: the negative spin solution possible only for realistically high inclinations and the positive spin 
solution for close to top view inclinations.  The allowed region of positive spin solutions shrinks with the black hole mass, giving tight constraints for the parameter space.}
\label{fig:kontur}
\end{figure*}

The computations were performed using a grid in the black holes mass, for the values from 
$10^9 M_{\odot}$ till $7 \times 10^9 M_{\odot}$, with the step of $10^9 M_{\odot}$. 
The broad band spectrum was 
satisfactorily fitted to the data for a broad range of the parameters, i.e. the errors for each of the 
measured quantities separately  are very large although the  photometric data are quite accurate. This reflects
considerable degeneracy in the fitted parameters. 
We illustrate it in Fig.~\ref{fig:kontur}. 

For every black hole mass between $2 \times 10^9 M_{\odot}$ and  $7 \times 10^9 M_{\odot}$ there is an extended
complex region with the acceptable $\chi^2$ in the  $\dot m$ and $a$ plane. For the black hole mass 
$2 \times 10^9 M_{\odot}$   the acceptable fit
is found only for negative spin of the black hole. The accretion rate is moderate, usually below 0.4, so the use of the
Novikov-Thorne model is acceptable. The inclination angle of the disk is small. For the black hole mass 
 $3 \times 10^9 M_{\odot}$ both negative and positive spin of the black hole are allowed but the best fit value
appears at the positive side, at $a \sim 0.3$, and the rotation faster than $a = 0.4$ is excluded for this mass. 
The allowed range of the inclination angles varies considerably along the allowed contour, and the large negative values
of the black hole spin require large inclinations, above 45 deg. For the black hole mass  $4 \times 10^9 M_{\odot}$
a qualitative change is visible: previously seen single elongated region breaks into two elongated regions, separated by the region with intermediate disk inclination and unacceptable  $\chi^2$ value. The high positive spin acceptable region 
of solutions is characterized by low values of the inclination angle, below 50 deg, while the region of negative spin solutions requires inclinations in excess of 60 deg. The trend deepens at still higher masses. 
For the black hole mass $7 \times 10^9 M_{\odot}$ the two regions are widely separated. The positive spin range allows only
for a high spin, between 0.8 and 0.998, and the inclination close to top view (less than $\sim 35$ deg). 
The negative spin branch requires very high inclination, above $\sim 75$ deg. There are no solutions outside this 
black hole mass range.

The parameter degeneracy is not unexpected.
The observed spectrum is characterized by the 
normalization in the optical range, the
bending at the logarithmic frequency $\sim 15.2$ and the position of the far-UV Galex point. 
In the model spectrum, the normalization depends on the mass, accretion rate and 
inclination angle of the source, the spectrum extension is determined by the mass, accretion rate and the Kerr
parameter, and the far-UV Galex point gives constraints on the viewing angle due to
the relativistic effects changing the spectral shape. The last constraint is very week if the disk does not extend
close to the black hole horizon. This is the reason why for negative spin the allowed region for a given mass is so broad
while it narrows for solution close to maximally rotating black hole. This is also the reason why the allowed region
for large mass consists of two parts: for negative spin (inner disk starting far from the black hole horizon) 
there are practically no constraints for the inclination and 
the rise of the accretion rate required to fit the spectrum extension does not overproduce the luminosity since the
rise in inclination can compensate for this effect. For fast positive disk rotation the far UV Galex point, 
although known with considerable error, limits the inclination very effectively due to the relativistic effects.

Black hole mass of $10^9 M_{\odot}$ does not fit the data points at all, even for counter-rotating disk. 
Low black hole mass requires high accretion rate to reproduce the source bolometric luminosity but high
accretion rate implies that the disk spectrum extends to too high frequencies in order to match the two Galex 
points. In principle, a successful model for such a small black hole can be found by assuming that the
accretion rate is large ($\dot m > 0.5$) but at the same time inner disk radius is much larger than  
implied by the extreme counter-rotating disk (i.e. much larger than $9 R_g$).
However, this would mean introducing another free parameter, and in addition, all current studies of accretion in AGN as 
well as in Galactic black holes indicate that the disk extends down to the ISCO for high accretion rate, and becomes likely
detached from ISCO when the accretion rate drops considerably below the Eddington value.

Black hole masses of $8 \times 10^9 M_{\odot}$ and higher allow only the counter-rotating solution with the inclination 
close to 90 deg. The high spin positive branch disappears since the accretion rate cannot drop with the rise of the black
hole mass if the spectrum extension has to be fitted (there is no space for further increase of the Kerr parameter beyond 
0.998) so the inclination angle decreases (otherwise the overall normalization is too high), and such a decrease leads
to the spectrum which is relativistically broadened too much to fit the Galex point. The solutions for extremely high 
inclination are unrealistic since the accretion disk itself has a final thickness (although the effect is not 
included in the current model). In addition, the view towards the quasar disk is expected to be shielded by 
the dusty/molecular torus,
(although the estimates of  a torus opening angle $\sim 40 - 45^{\circ}$ are available for Seyfert galaxies, e.g. 
Tovmassian 2001, Levenson et al. 2002), and the opening angle of the torus in bright quasars is likely to be larger 
(e.g. Arshakian 2005). Trend with luminosity is also supported by the fact that  
there are only few type 2 QSO (see e.g. Del Moro et al. 2009 for a recent detection), and in type 1 QSO 
only some 30 \% of the optical/UV bolometric luminosity is intercepted by dust and reemitted in the near-IR (e.g. Richards et al. 2006). The inclination
angle of $40^{\circ}$ corresponds to the inner radius located at $\sim 25 \, r_g$. Adopting larger value of the hardening factor would push
the inner radius still more out for the same inclination angle.

\begin{figure}
\epsfxsize=8.5cm
\epsfbox{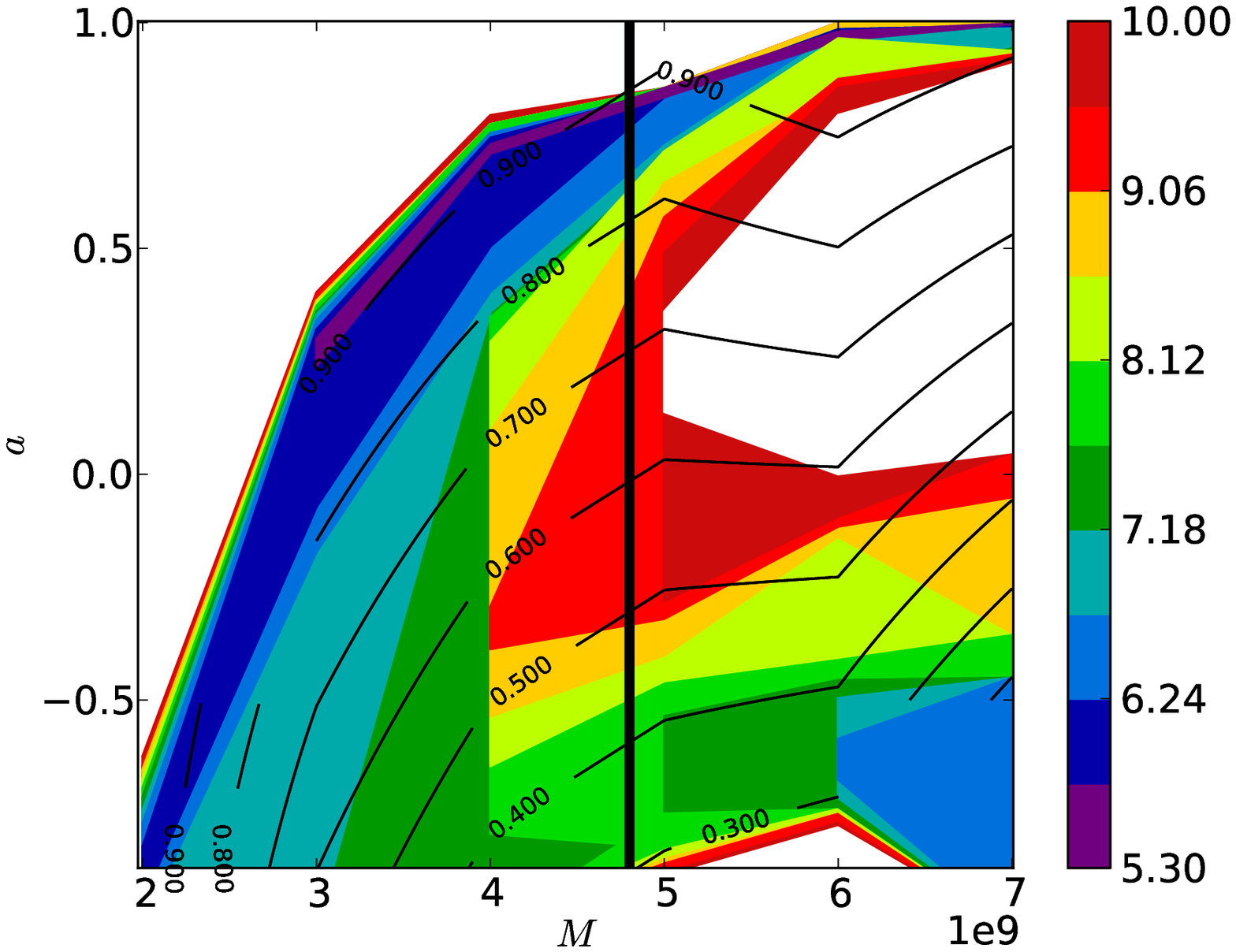}
\caption{The contour error for the Kerr parameter, $a$ for the black hole for the allowed range of the black
hole masses ({\bf the mass upper limit from Mg II line is marked with the vertical line}). }
\label{fig:kontur_total}
\end{figure}

Thus, if  there are no additional constraints for the black hole mass, 
the range of allowed values of the Kerr parameters covers a whole space from -1 up to 0.998. The summary of the fits
is shown in Fig.~\ref{fig:kontur_total}. We also mark there a line which divides the solutions with realistic inclination 
angles from those which are highly inclined. We adopted the dividing line somewhat arbitrarily at $\cos i = 0.7$ (i.e. $\sim$ 45 deg). 

In order to narrow the spin range we now use the constraints for the black hole mass which can be derived from the Mg II 
line width. Hryniewicz et al. (2010) give the line FWHM of $5490 \pm 220$ km s$^{-1}$ combined with the formula for the black hole mass of Kong et al. (2006)
\begin{equation}
M = 3.4 \times 10^6 ({\lambda L_{3000 \AA} \over 10^{44} {\rm erg ~s}^{-1}})^{0.57 \pm 0.12} ({FWHM ({\rm MgII}) \over 1000 {\rm erg ~s}^{-1}})^2 M_{\odot}
\end{equation}
leads to the value $M = 2.7 ^{+2.1}_{-1.2} \times 10^9 M_{\odot}$. The quoted errors are asymmetric since the (symmetric) error in the formula above is in the exponent. We mark those constraints in Fig.~\ref{fig:kontur_total}.
The maximum value of the black hole mass does not give any constraint for the lowest value of the black hole spin.
The minimum value of the black hole mass gives an upper limit for the black hole spin $a < 0.8$. The most likely value
of the black hole mass $ 3 \times 10^9 M_{\odot}$ constraints the black hole spin to the values $a < 0.4$, and if the
inclination is constrained to those of $\cos i < 0.7$ then the spin limit is $-0.6 <a < 0.4$, with the best fit value of
$a = 0.3$. Accretion rate in all those models is moderate, below 0.4 of the Eddington rate, therefore the assumption 
underlying the Novikov-Thorne model are satisfied. The summary of the fits are given in Table~\ref{tab:results}. 

\begin{figure}
\epsfxsize=8cm
\epsfbox{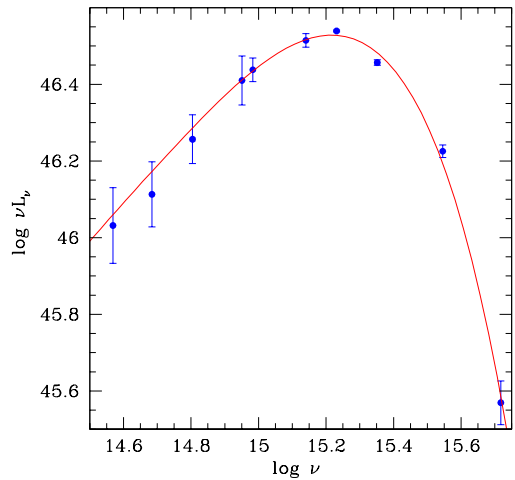}
\caption{The best fit of the Kerr black body disk with radius-dependent hardening factor to the spectrum, 
assuming $M = 3 \times 10^9 M_{\odot}$: $\dot m = 0.135$, $a = 0.3$, $ i = 0$ deg
(continuous line) and photometric points (dots with marked errors) 
for SDSS J094533.99+100950.1. 
}
\label{fig:best_fit}
\end{figure}

Exemplary fit to the photometric data points is shown in Fig.~\ref{fig:best_fit}. There is no systematic departure from
the data points in the whole energy range. The $\chi^2$ for this fit is 5.6 for 6 degrees of freedom.
The inclination of the disk in the best fit model is low ($i = 0^{\circ}$) which may seem surprising in view of the large Mg II line width. However, inclinations up to $40^{\circ}$ are allowed, if the black hole spin is zero, and up to $80^{\circ}$, if the spin is negative. 
On the other hand, although broad lines certainly have Keplerian
circular velocity component there is also some additional dispersion due to the outflow or inflow (see e.g. Shapovalova et al. 2004, Nikolajuk et al. 2006, Collin et al. 2006). In this case a considerable line width is expected even for a top view 
object. Supporting argument comes from the radio quiet quasar PG 1407+265 with a faint superluminal jet, where we have a viewing angle of a few degrees and still very broad lines of $\sim 10 000$ km s$^{-1}$ (Blundell et al. 2003). Best fit accretion rate of $0.135$ locates our source within the parameter range obtained 
for SDSS quasars by Kelly et al. (2010).

\begin{table*}
\caption{The best fits for the adopted disk models. Errors of the parameters are described in the text
since the contour errors are extremely elongated.}
\label{tab:results}
\begin{center}
\begin{tabular}[t]{lcccccccc}
\hline

Black hole mass          & a    &  $\dot m$ &  $\cos (i)$ \\
     &     &      \\
\hline
\\
$2 \times 10^9 - 7 \times 10^9 M_{\odot}$ {\rm (whole range)}  &  $0.3^{+0.7}_{-1.3}$ &  $0.14^{+0.26}_{-0.11}$    &  $1.0^{+0.0}_{-0.8}$ \\
$2 \times 10^9 - 5 \times 10^9 M_{\odot}$ {\rm (Kong constraint)} &  $0.3^{+0.5}_{-1.3}$ &  $0.14^{+0.26}_{-0.11}$    &  $1.0^{+0.0}_{-0.8}$ \\
$2 \times 10^9 - 5 \times 10^9 M_{\odot}$  &  $0.3^{+0.5}_{-1.3}$ &  $0.14^{+0.26}_{-0.11}$    &  {\rm above 0.7 (adopted)}\\
$3 \times 10^9$ {\rm (fixed)}  &  $0.3^{+0.1}_{-0.9}$ &  $0.14^{+0.11}_{-0.02}$    &  {\rm above 0.7 (adopted)} \\

\hline
\end{tabular}
\end{center}
\end{table*}

\section{Discussion}

Weak Line Quasar SDSS J094533.99+100950.1 is one of the few objects with exceptional properties of their broad 
emission lines. Usually strong CIV line is almost in-existent in this source, but the Mg II line and Fe continuum are
there, and Mg II line is broad, with FWHM of 5900 km s$^{-1}$. The aim of this paper was to study whether the continuum
properties are also exceptional and what constraints such a continuum imposes onto the state of accretion disk powering 
the black hole in this object.

As was shown by Hryniewicz et al. (2010) the IR/optical continuum of SDSS J094533.99+100950.1 is fairly typical 
for an SDSS quasar but the spectrum turns down in far-UV. Here we demonstrated that this broad band spectrum is 
well fitted by a simple accretion disk model. The decline in the far-UV is consistent with exponential and 
corresponds to the maximum temperature in the disk. No additional spectral component is needed, in particular in the far-UV.
 This source thus provided an exceptional opportunity for an attempt of spin determination
using the disk fitting method. 

When we combined the constraints for the black hole mass in SDSS J094533.99+100950.1 and we rejected the solutions which 
implied very high unrealistic inclination of the accretion disk in this system, we obtained that the
spin black hole spin is smaller than 0.8, and the negative spin values are also allowed. The broad spin range reflects
large error in the black hole mass determination as well as the considerable degeneracy in the parameter fits. 
The accretion rate for all acceptable models is in the range of 0.05 - 0.4 of the Eddington rate. This is fully
consistent with no departures from expectations of the standard Novikov-Thorne disk model.
 
Such a purely thermal disk-dominated state is not very frequent in active galactic nuclei, although
the disk emission 
explains the properties of the optical/UV continuum in most objects (e.g. Shields 1978, Malkan \& Sargent 1982, Czerny \& Elvis 1987, Sun \& Malkan 1989, Laor \& Netzer 1989, Koratkar \& Blaes 1999, Blaes et al. 2001, Blaes 2004, Bonning et al. 2007, Davis et al. 2007), including
the aspect of variability (e.g. Liu et al. 2008). 
The additional argument in the favor of the accretion disk interpretation of 
the optical/UV spectra came from the IR polarization measurements which uncovered
the blue disk component hidden in the dust emission.  
(Kishimoto et al. 2008).
Seyfert galaxies have usually
much redder optical slopes than predicted by the standard accretion disk (Thakur \& Sood 1981),  due to the contamination 
by the host galaxy as well as due to irradiation by the X-rays (e.g. Sergeev et al. 2005). 
Narrow Line Seyfert 1 galaxies show strong 
effect of Comptonization in the far-UV and soft
X-ray band. Quasar composites like the composite of Zhang et al. also show the broad optical-UV power law not well fitted
by a multicolor black body disk which implies the role of electron scattering or outflows. However, an older 
composite by Francis et al. (1991) made of exceptionally bright quasars show similar purely thermal shape, 
well fitted by a disk up to the Lyman $\alpha$ line (Koratkar \& Blaes 1999). Among the quasars studied by 
Richards et al. (2003), the bluest composite could be well 
approximated by a simple disk above $\log \nu = 14.7$ but there was a noticeable disagreement below that frequency, in the
red part of the spectrum (Czerny et al. 2004), possibly due to the host galaxy contamination.

In the case of accreting black holes in binary systems such pure thermal disk states are also not frequent. 
They usually appear when the source luminosity is somewhat lower that in the Very High State when
strong Comptonization (steep power law) is seen. Such a state is usually selected in order to determine the black 
hole spin, and the results are quite precise (Shafee et al. 2006, McClintock et al. 2006, 
Middleton et al. 2006, Gou et al. 2010) although there are some differences between the results obtained by various 
authors (e.g. $a = 0.7$ in Middleton et al. and $a > 0.98$ in McClintock et al. for GRS 1915+105).
However, in the case of our quasar we have no information on the inclination angle which is frequently 
available for binary 
systems thus reducing considerably the spin error.

The value of the black hole spin in SDSS J094533.99+100950.1 obtained from the basic model is comparable to the
values derived for other AGN using different methods. Higher value of  the spin is
implied by the simple analysis of the iron line profile in MCG -6-15-30 ($a = 0.99$, Brenneman \& Reynolds 2006, $a = 0.95$, 
Goosmann et al. 2006), although the inclusion of the disk reflection 
below the ISCO can decrease this spin significantly (Reynolds \& Begelman 1997; however see Reynolds \& Fabian 2008 for counter-argument). 
Spins in other AGN also obtained from the iron line shape are lower: $0.6 \pm 0.07$ in Fairall 9 (Schmoll et al. 2009), $0.6 \pm 0.2$ in AGN SWIFT J2127.4+5654 (Miniutti et al. 2009).
The accretion efficiency $\eta$ implied by our spin determination for SDSS J094533.99+100950.1 is below 0.12, 
rather consistent with determinations based 
on Soltan (1982) 
approach or quasar clustering properties ($\eta = 0.05 - 0.1$, Shankar et al. 2009; $\eta > 0.17$, Shankar et al. 2010; $0.18$ 
for AGN with black hole mass above $10^9 M_{\odot}$, Cao \& Li 2008).

\subsection{dependence on model assumptions}

\subsubsection{the role of advection}

\begin{figure}
\epsfxsize=8.5cm
\epsfbox{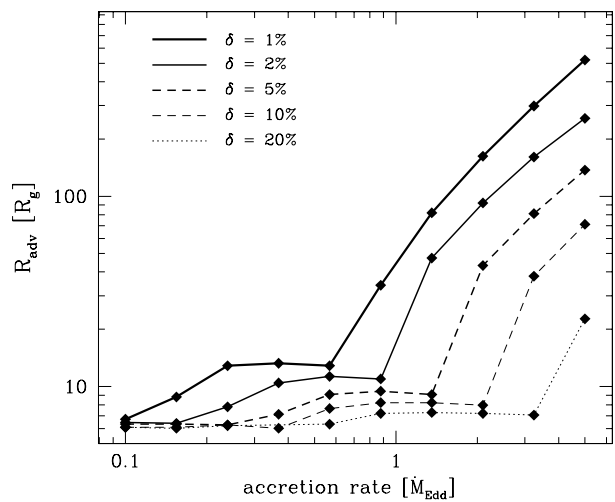}
\caption{The curves show the radii where the slim disk effective temperature differs by 1 to 20 per cent, 
correspondingly, from
the effective temperature of the Novikov-Thorne model, as a function of accretion rate.
}
\label{fig:advec}
\end{figure}

Since we used the Novikov-Thorne disk model which neglects the effect of advection we performed tests of the 
role of the advection term and transonic character of the flow which in principle should be 
included in the disk equations (Muchotrzeb \& Paczynski 1982, 
Abramowicz et al. 1988, S\c adowski 2009). We compared the effective temperature profile in Novikov-Thorne disk with
the effective temperature profile in the slim disk, using the code described in detail
by S\c adowski (2009), assuming the viscosity parameter 0.01. 
At larger radii the effect of advection is weak and increases inwards, and the overall effect increases with
the rise of the accretion rate.
Fig.~\ref{fig:advec} shows the radii where the slim disk temperature differs by a fixed fraction for a given
accretion rate. We see that for the accretion rate $\dot m = 2$ the disk temperature is modified by 5 per cent 
or more for radii smaller than $50 \, r_g$, and for accretion rate $\dot m = 4$  the affected region 
increases up to $100 \, r_g$. Below the Eddington limit the temperature at only the innermost few $r_g$ is 
slightly  modified. Kulkarni et al. (2011) also concluded that the Noviko-Thorne model applies very well up to 
$\dot m \approx 0.3$, and up to values $\dot m \approx 0.7$ the errors are not very large.

The second effect of the slim disk is the considerable disk thickness, so the photon emission cannot be approximated
as taking place in the equatorial plane. However, again this effect is most important in the innermost
part of the disk. For test purposes, we repeated the fits of the slim disk model with finite disk thickness
for the case of isotropic emission, hardening factor 1.4, accretion rate $\dot m = 2.11$ and the inner radius
cut at 60 $r_g$. The quality of the fit was the same as that of Novikov-Thorne model with neglected disk
thickness, and the best fit inclination angle was very similar in both cases: 77.91$^{\circ}$ vs. 78.09$^{\circ}$.

\begin{table*}
\caption{The best fits for non-standard models, for single value of the black hole mass (see text).}
\label{tab:results2}
\begin{center}
\begin{tabular}[t]{lcccccccc}
\hline

Model         &   $E(B-V)_{intrinsic}$ &  $f_{hard}$   & M  & a    &  $\dot m$ &  $\cos (i)$ & $\chi^2 $\\
     &     &      \\
\hline
\\
 Kerr bb (positive spin)  & 0 & 1  &$3.0 \times 10^9 M_{\odot}$  &$0.6^{+0.1}_{-0.3}$ &  $0.11^{+0.04}_{-0.02}$    &  $0.9^{+0.1}_{-0.1}$       & 6.6 (7 d.o.f.) \\
 Kerr bb limb bright.& 0 &  1   & $3.0 \times 10^9 M_{\odot}$&$0.8^{+0.1}_{-0.4}$ &  $0.06^{+0.04}_{-0.01}$    &  $1.0^{+0.0}_{-0.2}$      & 6.5  (7 d.o.f.)\\
 Kerr bb limb dark.  & 0 &  1   & $3.0 \times 10^9 M_{\odot}$& $0.0^{+0.1}_{-1.0}$ &  $0.31^{+0.12}_{-0.15}$    &  $1.0^{+0.0}_{-0.5}$       & 6.4 (7 d.o.f.)\\
 Kerr bb             & 0.1 (Gaskell) & 1 & $2.7 \times 10^9 M_{\odot}$&$0.0^{+0.16}_{-0}$ &  $0.34^{+0.01}_{-0.03}$    &  $0.97^{+0.03}_{-0.02}$       & 18.8 (7 d.o.f.) \\   
 slim bb            & 0.1 (SMC) & 1 & $2.7 \times 10^9 M_{\odot}$&$ > 0.3$ &  $ 0.90^{+1.10}_{-0.63} $    &  $0.20^{+0.33}_{-0.09}$       & 2.4 (7 d.o.f.) \\ 
\hline
\end{tabular}
\end{center}
\end{table*}

The results quoted above depend on the underlying assumptions, i.e. black body isotropic emission and no 
intrinsic reddening of the quasar spectrum by circumnuclear dust. We tested the sensitivity to those assumptions by performing 
additional fits. 

\subsubsection{hardening factor}

In the fits presented in Section~\ref{sect:results} we used the hardening factor formula described in Sect.~\ref{sect:hardening}.
 In principle, this
freedom can be reduced by performing proper radiative transfer in the disk atmosphere. 
Several non-black body disk spectra models were actually developed but the satisfactory tables for those
factors are not available yet. Hubeny et al. (2000) considered only 
hydrogen-helium atmosphere.  Davis et al. (2007) applied the  such hydrogen-helium
models in a statistical approach to quasar slopes in SDSS sample. However, the negligence of the metals is not acceptable
in case of relatively cool AGN disks.
 Complete  models with numerous bound-free transitions were developed by Hubeny et al. (2001), used by 
Davis et al. (2005) and the models based on this approach were even implemented in {\sc xpec} 
to fit the spin in black hole binaries (see Gou et al. 2010 and the references therein) but the tables do not cover 
supermassive black holes. 
AGN spectra were computed by Rozanska \& Madej (2007) but only for a single value of the accretion rate, 
$\dot m = 0.03$, and they included only hydrogen and iron transitions (lines and continua). 
Models by Garcia \& Kallman (2010)  include recent atomic data but assume strong X-ray irradiation
and constant density so they display strong emission edges starting from 1.5 eV, and they were calculated at a single radius
for a very low Eddington ratio ($1.6 \times 10^{-3}$). The computations of the radiative transfer may additionally be complicated by the dissipation close to the disk surface although Davis et al. (2009) argued that a strong effect due to the  
presence of the magnetic field in the outer disk layers is not expected.  

However, the value of the hardening factor is not very high in AGN due to the relatively low disk temperature, unless the disk surface is strongly
irradiated. We tested the sensitivity of the present results by comparing the results for our hardening factor to the results for a black body
emission. We repeated the computations for the fixed black hole mass value of $3 \times 10^9 M_{\odot}$. The solution has the properties of the solutions
with hardening factor but for slightly higer mass, so two separate branches appear, one branch for positive spin and low inclinations, 
and the second unphysical branch with large negative values of the black hole spin and high inclination angles. We give the results in 
Table~\ref{tab:results2} for positive spin branch only. The allowed range of the spin is moved towards higher values, and the accretion rate 
is limited to lower values.

In principle, it is also possible to find the constraints for the hardening factor directly from the data
since the radius-dependent 
correction leads to the departure of the spectrum from the standard multicolor black body. This should be most easily seen in the 
rest frame optical part, but the broad band spectrum there is only covered by the 2MASS points with large error bars.

\subsubsection{anisotropic emission}

In our basic model we assumed that the local emission is isotropic.  
However, the disk spectra may show limb-darkening or limb-brightening effects. The models
discussed in the literature are not conclusive on this issue. Models developed for
small black hole masses appropriate for galactic black holes imply the scattering-dominated
limb-brightened atmosphere (e.g. Davis et al. 2005). Irradiated models of AGN disks in X-ray 
band also imply considerable limb-brightening (e.g. Zycki \& Czerny 1994, Rozanska \& Madej 2008)
but the optical/UV part of the spectrum shows rather limb-darkening trend (Rozanska \& Madej 2008)
since the metals provide considerable opacity at temperatures of $\sim 10^4 - 10^5$ K. 

We parameterize the effect in a simple way assuming that the local emissivity is described by
\begin{equation}
I = I_0 {2 \over 2 + \xi_{limb}}(1 + \xi_{limb} \cos \theta),
\end{equation}
where $\theta$ is the local angle from the normal to the emissivity surface and $I_0$ is the isotropic
intensity. For example, the 
scattering-dominated atmosphere is approximately described by $\xi_{limb} = 1.8$, while the 
spectra of Rozanska \& Madej (2008) showed limb-darkening with $\xi_{limb}$ reaching -2/3 in far-UV.

We calculated the models for exemplary limb-brightening and limb-darkening laws, otherwise just assuming the local black 
body emission. We performed computations again for a fixed black hole mass. 
The assumption of the limb-brightening moves the allowed range of solutions towards much larger black hole spin, while the
limb-darkening decreases allowed spin considerably. 

\subsubsection{reddening by dust and slim disk models}

The photometric data points described in Sect.~\ref{sect:data} were corrected for the 
Galactic reddening. However, intrinsic reddening can also affect quasar spectra 
(e.g. Richards et al. 2003, and the references therein). The turn-off of the SDSS  
J094533.99+100950.1 in the far-UV band can
in principle be due to the dust. Therefore, we test this effect by allowing for some 
arbitrary amount of
circumnuclear dust along the line of sight. Since the dust properties are uncertain, 
and likely evolve with the
redshift (see e.g. Gallerani et al. 2010 for a recent paper on this subject) we
 consider three extinction curves
likely to represent the quasar dust properties. The first one is the SMC extinction curve
broadly used for AGN (e.g. Richards et al. 2003, Davis et al. 2007) due to its lack of 
2200 \AA~ feature, also absent in quasars.
We specifically applied the SMC bar curve AzV 18 of Cartledge et al. (2005). 
The second curve comes from Gaskell et al. (2004), and the third one from Czerny et al. (2004). 

Since the determination of the extinction curve is also biased with some errors, we 
calculated the error bars of the
dereddened photometric points by combining the photometric error bars with the 
errors in extinction curve parameters.
In the case of the SMC curve described with six parameters with error bars 
(Cartledge et al. 2005) we performed Monte Carlo 
simulations of the errors assuming Gaussian distribution of the extinction curve 
parameters around their corresponding 
best fit values, calculating the dereddened spectrum for every realization of the
 extinction curve and we determined
the mean and the error of each photometry point due to extinction from the 
resulting distribution.

We fixed the amount of reddening arbitrarily at rather high value of $E(B-V) = 0.1$ 
in order to see the effect more clearly. 

\begin{figure}
\epsfxsize=8.5cm
\epsfbox{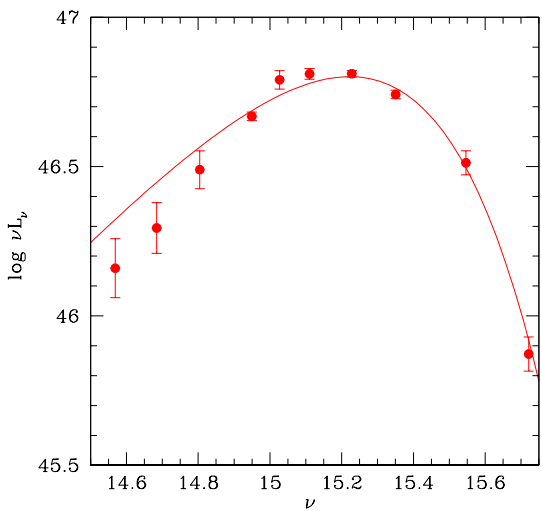}
\caption{The broad band spectrum of SDSS  J094533.99+100950.1 dereddened by $E(B-V) = 0.1$ with the Gaskell extinction curve
and the best fit of the disk model (fit not acceptable, with $\chi^2 = 18.8$ for 7 d.o.f). 
}
\label{fig:spec_gaskell}
\end{figure}

The significant amount of reddening combined with the Gaskell extinction curve leads to the continuum shape
which is not well modeled by the disk. The optical part of the dereddened continuum is too steep 
(see Fig.~\ref{fig:spec_gaskell}) although the
remaining UV part still drops roughly exponentially and it is well followed by the thermal spectrum. Assuming disk disruption
we can obtain solutions with larger accretion rates and inclination angles than given in the Table~\ref{tab:results} but the fit quality
does not improve. Acceptable fits can be obtained only by introducing another free parameter to the model in the form of the
small outer disk radius $250 r_g$. 

The application of the SMC extinction curve leads to the spectra which are well fitted by the disk models. The IR/optical spectrum
is relatively  unaffected but the position of the maximum on the $\nu L_{\nu}$ diagram shifts to the right so much higher accretion rates
are indicated. 

Since test computations in this case implied accretion rate above the Eddington limit, we performed computations
assuming  the  slim disk model and using the code of S\c adowski (2009). 

In this case the disk structure was not 
specifically calculated for
the required accretion rate and the Kerr parameter but a set of pre-defined 
table models (for the detailed
description of the code, see S\c adowski 2009) were used. The tables were computed 
for three
values of the Kerr parameter (0, 0.3, 0.998) and 20 values of accretion rate, 
spaced logarithmically. The viscosity parameter $\alpha $ was fixed at the value of 0.01. 
A value of this order is favored by the studies of the quasar variability in the 
optical band (e.g. Siemiginowska \& Czerny 1989, 
Starling et al. 2004) as well as by the simulations of the magneto-rotational 
instability (e.g. Hirose et al. 2009).
The disk thickness was stored and the ray tracing was done starting 
from the disk surface instead 
from the equatorial plane. Self-irradiation of the disk, however, was not 
included, and the photons returning to the disk were considered as being lost.

\begin{figure}
\epsfxsize=8.5cm
\epsfbox{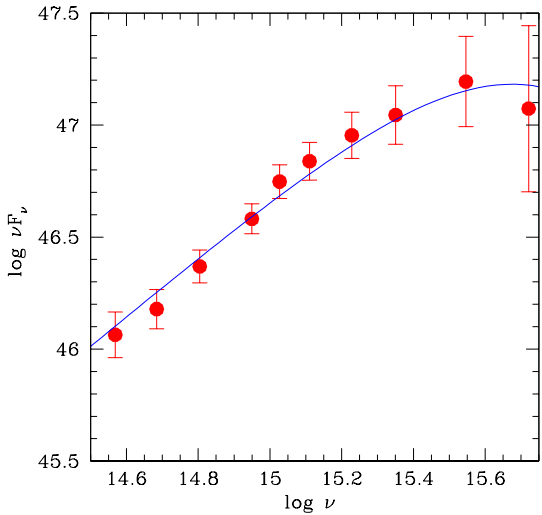}
\caption{The broad band spectrum of SDSS  J094533.99+100950.1 dereddened by $E(B-V) = 0.1$ with the SMC extinction curve
and the exemplary fit of the slim disk model: $M = 2.7 \times 10^9 M_{\odot}$, $a = 0.998$, $\dot m = 0.6$, $\cos i = 0.28$. 
}
\label{fig:slim_spec}
\end{figure}

The solution for the 
rotating black hole was favored by the data, with the best fit value 0.998 (the highest trial value). The fit quality is formally better, but it is due to the fact that the error bars strongly increased in the  UV part due to the errors in the extinction law. The implied inclination was also high.  This high inclination is imposed by the fact that the peak of the spectrum on $\nu L_{\nu}$ diagram is now
shifted towards higher frequencies which implies higher accretion rate while relatively unchanged normalization of the spectrum 
in the IR/optical band is inconsistent with the higher accretion rate unless higher inclination compensates for this effect. 
Intermediate amount of reddening (e.g. $E(B-V) = 0.03$, as on average seen in SDSS quasars, Hopkins et al. 2004) will lead to lower accretion rates and lower inclination than implied by the
solution for $E(B-V) = 0.1$. Similar results are obtained if the extinction curve of Czerny et al. (2004) is used.

We have not tested more complex possibilities, like the crystalline dust recently considered in the context
of quasar Ton 34 by Binetter \& Krongold (2008), since for our objects we have only photometric points in the far UV band.

Large inclinations obtained in the fits indicate that strong reddening is rather unlikely in this source. Independent argument against the
significant reddening comes from the paper by Punsly \& Zhang (2011). They have found at the basis of the large sample of AGN 
that Mg II line can be used as a bolometer. The measured equivalent width of the Mg II line in SDSS J094533.99+100950.1 corresponds to the
expected bolometric luminosity of the source $6.4 \times 10^{46}$ erg s$^{-1}$, while the bolometric luminosity obtained directly by integrating the broad band spectrum gives the comparable value, $6.2 \times 10^{46} $ erg s$^{-1}$.

We stress that we tested stationary accretion disk models although Hryniewicz et al. (2010) postulated that the quasar activity has just started 
in this object. However, it is still possible that the disk has just achieved a stationary state. In time-dependent disk  computations we always see that 
the disk during outburst is well approximated by a stationary model (but this is not true between the outbursts when the mass accumulates in the disk). 
The same is seen observationally, for example from mapping the outbursts of cataclysmic variables (Horne 1985; see Halevin et al. 2010 for a recent application).

The disk properties: pure disk state, no evidence of X-ray emission and the disk extending down to ISCO is consistent with the freshly started activity episode, although it does not prove it. In Galactic sources, the outbursts start as X-ray dominated states and a thermal disk state (disk-dominated Soft State or Compton-dominated Very High State) appears later (Done et al. 2007). 

The global evolution of SDSS J094533.99+100950.1 seems rather advanced. Young, high redshift quasars ($z > 6$ are accreting close (or just above) the Eddington limit, the FMHW of their Mg II
line is relatively small, $\sim 3000$ km s$^{-1}$, and the kinematical width is correlated with the UV luminosity 
(Willott et al. 2010). If this correlation is applied to our object we would expect the luminosity above $10^{47}$ erg s$^{-1}$,
larger than the observed luminosity ($6.2 \times 10^{46}$ erg s$^{-1}$). The black hole mass of
SDSS J094533.99+100950.1 is also an order of magnitude higher than the black hole masses of Willott et al. sample. Therefore, 
the object seems to be more evolved, and the present reactivation, if confirmed, would be one of the many episodes.

\subsection{the influence of the continuum on the emission line properties}

We fitted the emission line properties of the quasar using the best fit 
continuum from the Sect.~\ref{sect:results}. The results are given in 
Table~\ref{tab:lines_bb_nolimb}. We see that the emission line properties did
not change much in comparison to those derived using a power law quasars continuum
(Table~4 in Hryniewicz et al. 2010, first three lines). 
The Mg II line is now even slightly broader and
stronger, C III] line is almost invisible, only the C IV line is now more significant,
with the equivalent width of $5.0$ \AA~ instead of 1.5 \AA, and its kinematical width is
comparable to the kinematical width of Mg II. However, such a C IV strength
is still much below the typical values for quasars, supporting the classification of 
the object as WLQ.

\begin{table*}
\caption{Emission line properties of
SDSS J094533.99+100950.1 for a disk underlying continuum from Fig.~\ref{fig:best_fit}.
}
\begin{tabular}{rcccccc}
\hline
\hline
Line & $\lambda_{0}$ & REW & FWHM & z\\
& [\AA]  & [\AA]   & [km s$^{-1}$]  &\\
\hline
Mg II $\lambda2800$ & $2800.68 \pm 0.78$ & $18.1^{+1.2}_{-1.2}$ &
$5910\pm 220$ & 1.66225\\
C III] + Si III] $\lambda1909$& $1905.4 \pm 3.1$ &
$0.83^{+0.55}_{-0.41}$ & $3700
\pm 3700$ & 1.65661\\
C IV  $\lambda1549$ & $1541.9 \pm 3.0$ & $5.0^{+2.4}_{-2.0}$ & $5900\pm
1400$ & 1.6494\\
\hline
\end{tabular}
\label{tab:lines_bb_nolimb}
\end{table*}

\begin{figure*}
\epsfxsize=18cm
\epsfbox{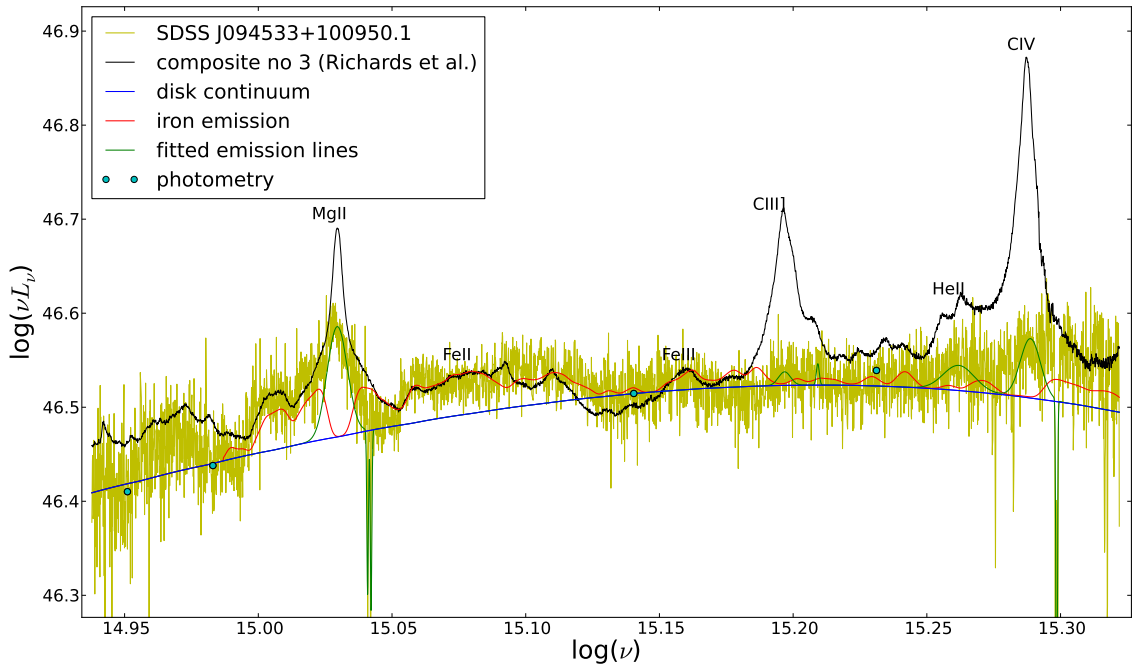}
\caption{The spectroscopic SDSS data for SDSS J094533.99+100950.1, with the continuum from Fig.~\ref{fig:best_fit}
(red line). 
Best iron line contribution fit is marked with magenta line, and the emission lines of Mg II, C III] and C IV
are marked with a green line. Photometric points used for fitting continuum are marked with a circles.
}
\label{fig:new_lines}
\end{figure*}

We do not iterate between the disk continuum fit and the Fe II and line contributions so we do not provide the
quantitative estimate of the fit of the spectroscopic part of the spectrum. Overall, the spectroscopy is well
reproduced by the sum of the components. However, there is a disagreement at some wavelengths. There is an
apparent excess between the $z$ and the next photometric point, as well as at the shorter wavelength to CIV. 
In the first region there may be a contamination by Fe II, but the Fe II template of Vestergaard \& Wilkes 
(2001) used in this 
paper does not cover
that range, and in addition the spectrum may be affected by inaccurate subtraction of the OH 
{\bf emission} features from the Earth atmosphere. In the second region the Fe II contribution seems too low although it
fits well the region close to Mg II. However, it is unlikely that the true continuum is harder since 
it already overpredicts $u$ photometry.

\section{Conclusions}

The broad band spectrum of the weak line quasar SDSS J094533.99+100950.1 is well fitted by the standard Novikov-Thorne disk model,
without any need for additional Comptonization in the far UV band. The best fit value of the accretion rate, 0.13, is much 
lower than the accretion rate in another Weak Line Quasar, modeled in detail (PHL 1811, $\dot m = 0.9$, Leighly et al. 2007a,
assuming non-rotating black hole)
which is consistent with narrow H$\beta$ line in PHL 1811 (1943 km s$^{-1}$, Leighly et al. 2007b) and 
broad Mg II in SDSS J094533.99+100950.1 (5910 km s$^{-1}$, this paper).
The maximum of the disk spectrum is well visible but nevertheless
fitting all the four parameters: black hole mass, accretion rate, spin and the disk inclination still leads to parameter degeneracy
since the errors in the far-UV Galex points are large. All spin values are allowed if no additional constraints are imposed.
If the limits for the black hole mass from the Mg II line are adopted, the black hole spin is constrained to values lower than 
0.8. The large errors  of the IR data points prevent from reliable testing the adopted radius-dependent hardening factor correction
to the local black body emission. Further theoretical work towards establishing proper hardening factors as well as the possible 
departure from the isotropic emission are essential for further progress of the spin determination by the disk fitting method.  

\section*{Acknowledgments}
We are grateful to the referee for suggestions which helped us to improve our paper significantly, and we thank Piotr \.Zycki for very helpful discussions. This work was supported in
part by grant NN 203 380136.

\ \\
This paper has been processed by the authors using the Blackwell
Scientific Publications \LaTeX\ style file.
\end{document}